# Scanning Microwave Microscopy of Aluminum CMOS Interconnect Lines Buried in Oxide and Water

Xin Jin, Kuanchen Xiong, Roderick Marstell, Nicholas C. Strandwitz, and James C. M. Hwang
Lehigh University
Bethlehem, PA 18015, USA
xij215@lehigh.edu

Marco Farina
Dipartimento di Ingegneria dell'Informazione
Università Politecnica delle Marche
Ancona 60131, Italy

Alexander Göritz, Matthias Wietstruck, Mehmet Kaynak
IHP, Im Technologiepark 25, 15236 Frankfurt (Oder), Germany

***Abstract*—Using a scanning microwave microscope, we imaged in water aluminum interconnect lines buried in aluminum and silicon oxides fabricated through a state-of-the-art 0.13-µm SiGe BiCMOS process. The results were compared with that obtained by using atomic force microscopy both in air and water. It was found the images in water was degraded by only approximately 60% from that in air.**

## I. Introduction

Scanning microwave microscopy (SMM) is an emerging scanning-probe technique for investigating nanostructured materials and devices with atomic-level resolution [1]. Unlike other scanning-probe techniques such as atomic-force microscopy (AFM) and scanning-tunneling microscopy (STM), the microwave signal of SMM can penetrate below the surface to probe buried nanostructures [2]. Further, because the energy of the microwave photon is on the order of µeV, SMM is particularly suitable for noninvasive probing of biological samples such as cells [3], bacteria [4], and subcellular structures [5]. However, biological samples must be kept in aqueous solution to stay alive and the microwave signal tends to be absorbed and scattered by water. Water also increases the drag on the probe. To begin to address such a challenge, this paper compares SMM images obtained on aluminum CMOS interconnect lines both in water and air and, in turn, compares them with that obtained by AFM. In all cases, comparisons were made between bare metal and metal covered by 20-nm of $Al_2O_3$ to assess the SMM's capability for probing buried nanostructures.

## II. Experimental Setup

The present samples comprise aluminum interconnect lines of a state-of-the-art 0.13-µm SiGe BiCMOS technology by IHP [6]. With this technology, metal surface is without native oxide. They were fabricated on a 200-mm high-resistivity (10 kΩ·cm) Si wafer then diced into 25 mm × 15 mm chips. Aluminum interconnect lines approximately 80-µm wide are embedded in approximately 1.3-µm thick $SiO_2$ chemo-mechanically polished to provide a good contrast on a relatively flat surface for SMM scanning. Between interconnect lines is an approximately 20-µm wide $SiO_2$ gap. Additionally, to provide a contrast between bare metal and metal buried under oxide, some chips were coated by atomic layer deposition (ALD) at 300°C of 20-nm $Al_2O_3$. The dielectric constants of $SiO_2$ and $Al_2O_3$ are measured to be approximately 4 and 7, respectively, by an impedance analyzer. Chips with bare metal and $Al_2O_3$-covered metal were mounted on the same piezoelectric stage with nanometer *x*, *y*, and *z* control for SMM and AFM scanning. Typically, information on topography, deflection/friction (magnitude/ phase in tapping mode), and microwave reflection were obtained simultaneously during the same scan.

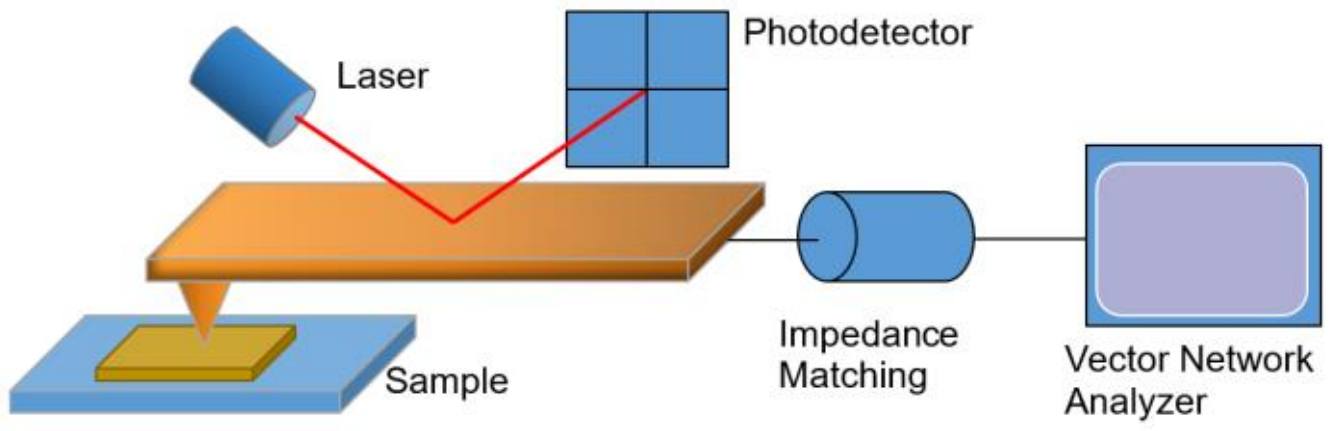

Fig. 1. Schematic of the present SMM setup.

Fig. 1 shows that the present SMM was based on a modified AFM [7] with the microwave signal channeled to a sharp tip for near-field interaction with the sample over an extremely small volume. A Rocky Mountain 12Pt400A full-metal scanning probe with smaller than 20-nm tip radius was used. Its spring constant is 0.3N/m, which is soft enough for contact-mode scanning. The microwave signal was provided by a 10 MHz–20 GHz Agilent N5230A PNA-L vector network analyzer (VNA) through an impedance matching network comprising coaxial cables and a shunt resistor. The reflected microwave signal, in terms of amplitude and phase of the return loss $S_{11}$, was also measured by the VNA after a 30-dB preamplifier. The AFM/SMM scanning speed was limited to 65 µm/s by the intermediate-frequency (IF) bandwidth of the VNA.

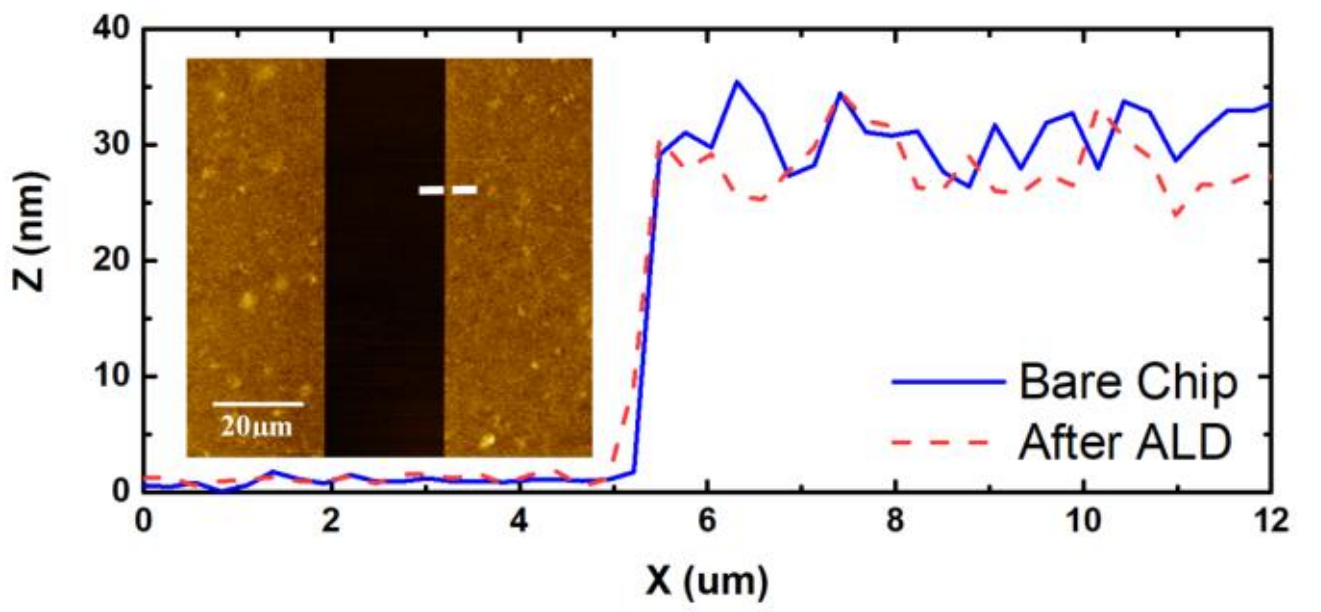

Fig. 2. AFM height profile across bare metal (—) and $Al_2O_3$-covered metal (- - -) embedded in $SiO_2$. Inset shows where the height profile was taken.

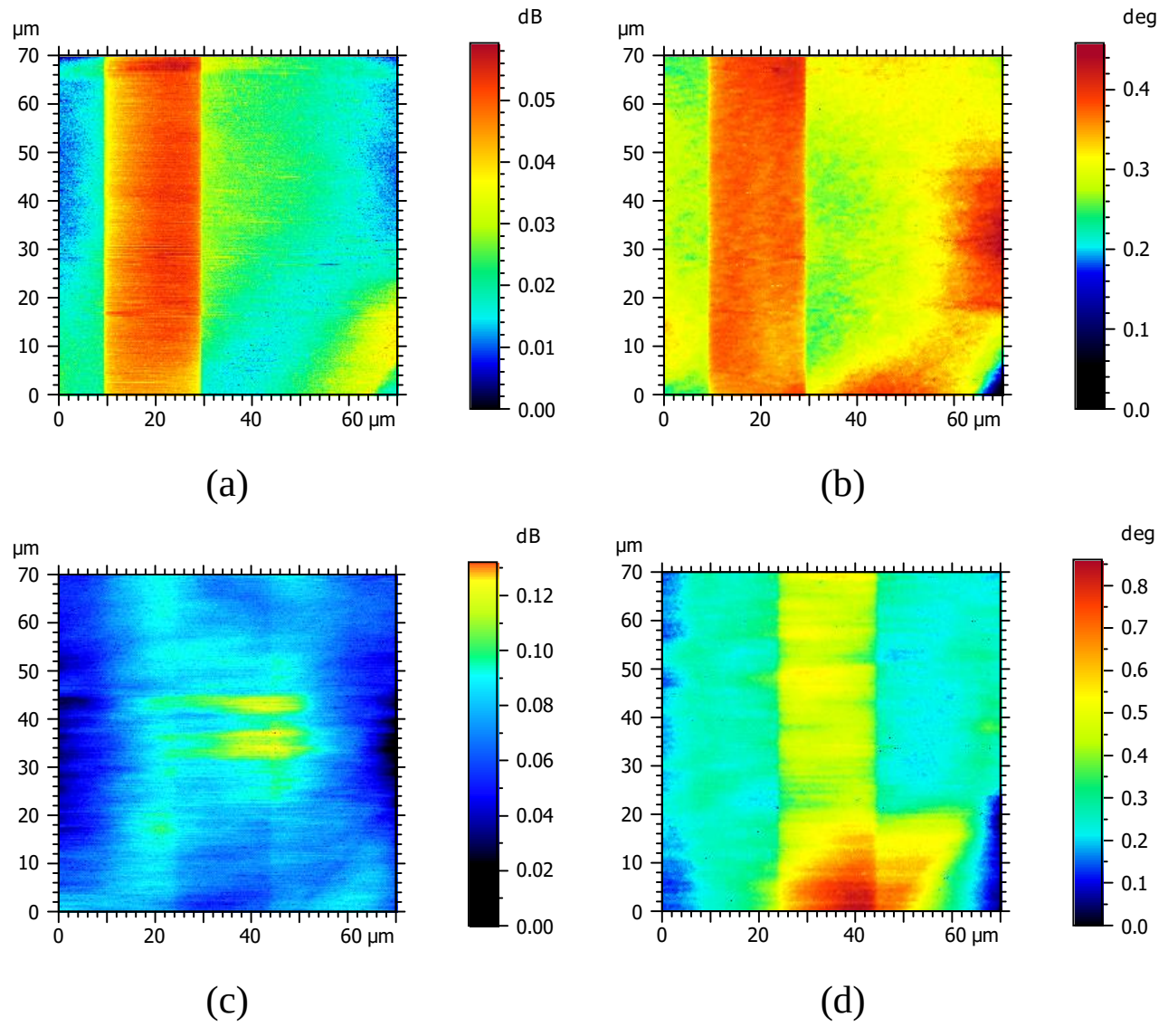

Fig. 3. SMM (a) amplitude and (b) phase images of bare metal vs. that of (c) amplitude and (d) phase of $Al_2O_3$-covered metal, with both samples in air.

## III. Results and Discussion

Using AFM, Fig. 2 shows that the chips with bare metal and $Al_2O_3$-covered metal have nearly identical topography due to the uniform and conforming characteristics of ALD, which minimizes the so-called "topography cross-talk" in SMM. Otherwise, topographic information can be coupled into the microwave return loss through the stray capacitance of the probe body. It can be seen from Fig. 2 that, on both chips, the metal rises approximately 30 nm above $SiO_2$ at a slope of approximately 12°. No attempt was made with smaller scanning area or correction for probe-sample convolution, since presently the exact topography is not of interest so long as it is identical.

Fig. 3 compares SMM images of bare metal and $Al_2O_3$-covered metal, both in air, obtained during the same AFM scan of Fig. 2. The microwave frequency was 9.03 GHz. The scanned area was 70 µm × 70 µm with 256 × 256 sampling points. The scan speed was 28 µm/s. It can be seen that the metal lines are discernable, although with lower contrast and higher noise, under SMM even when uniformly covered by $Al_2O_3$ and without the topographical information coupled in. Further, the degradation in contrast and noise can be used to determine the thickness of the uniform cover.

When the probe is immersed in deionized water, the resonance between the probe and the shunt resistor of the impedance-matching network shifts to a lower frequency (Fig. 4.), even with the probe far away from the sample. Therefore, a different frequency has to be used to maximize the sensitivity of the return loss when the probe is immersed in water while touching the sample.

Fig. 5 shows the AFM topography and deflection, and SMM magnitude and phase obtained simultaneously with the $Al_2O_3$-covered metal immersed in water. In this case, the SMM frequency is 8.72 GHz, while the scanning area and speed remain the same 28 µm/s. It can be seen that metal lines are clearly discernable under AFM and SMM. To quantify the degradation from air to water, Fig. 5(e) plots the normalized SMM phase change across the sample, with the average phases over metal and $SiO_2$ set to 0 and 1, respectively. It can be seen that the slopes at the metal edge are 0.636 and 0.270 in air and water, respectively. Thus, the degradation in image quality by water can be quantified to be approximately 60%.

## IV. Conclusion

SMM images were successfully obtained for aluminum CMOS interconnect lines buried in oxide and water. Compared to the images obtained with the same sample in air, the image quality was degraded by only approximately 60%. This shows that SMM can potentially be used to image live biological samples in aqueous solution with nanometer resolution.

## Acknowledgment

The project depicted is sponsored in part by the U.S. Department of Defense, Defense Threat Reduction Agency under Contract No. HDTRA1-12-1-0007 and Army Research Office under Contract No. W911NF-14-1-0665, as well as Research by the US Army Research Laboratory under Grant Number W911NF-17-1-0090.

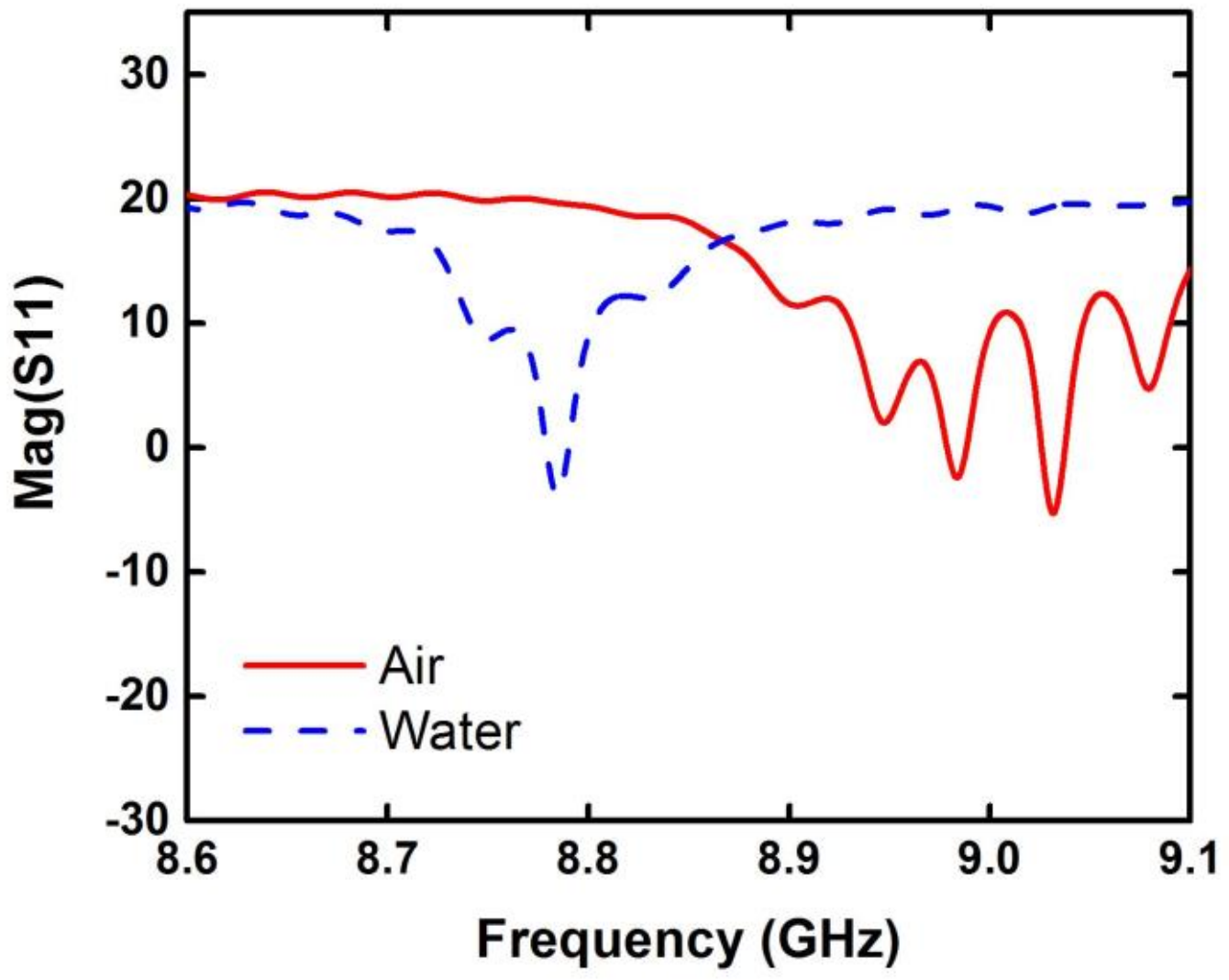

Fig. 4. SMM return loss across 8.60–9.10 GHz in both air (—) and water (- - -) without the probe touching any sample.

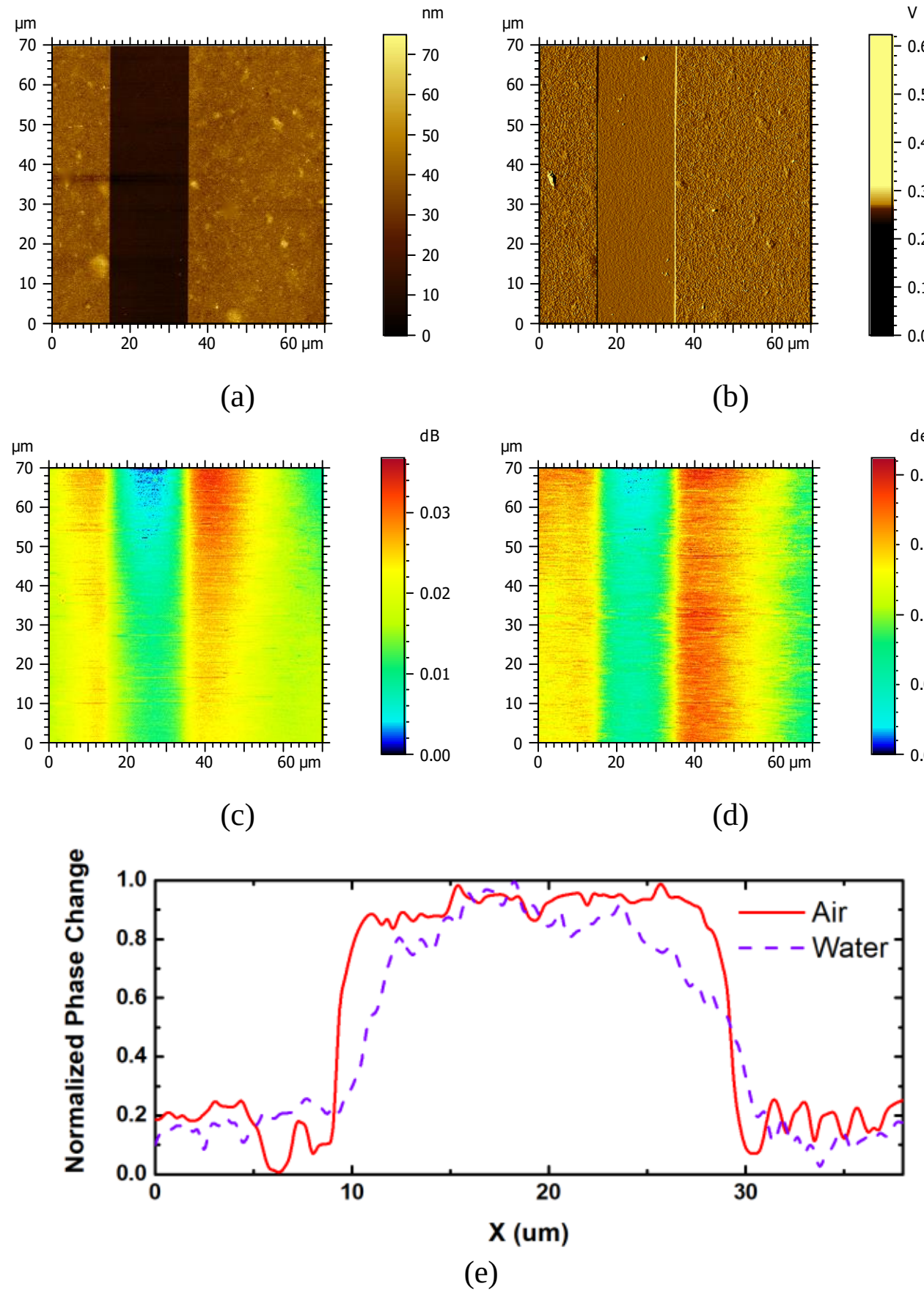

Fig. 5. (a) AFM topography, (b) AFM deflection, (c) SMM amplitude, and (d) SMM phase of $Al_2O_3$-covered metal in water. (e) Normalized SMM phase of the same sample in water and air.